\documentclass[preprint,showpacs,preprintnumbers,amsmath,amssymb]{revtex4}

\usepackage{epsfig}
\usepackage{graphicx}
\usepackage{dcolumn}
\usepackage{bm}

\newcommand{\ord}{{\cal O}}
\def\beq{\begin{equation}}
\def\eeq#1{\label{#1}\end{equation}}
\def\eeqn{\end{equation}}
\newcommand\iden{\leavevmode\hbox{\small1\normalsize\kern-.33em1}}

\let\jnfont=\rm
\def\RMP#1,{{\jnfont Rev.\ Mod.\ Phys }{\bf #1},}
\def\NPB#1,{{\jnfont Nucl.\ Phys.\ B }{\bf #1},}
\def\PLB#1,{{\jnfont Phys.\ Lett.\ B }{\bf #1},}
\def\EPJC#1,{{\jnfont Eur.\ Phys.\ Jour.\ C }{\bf #1},}
\def\PRD#1,{{\jnfont Phys.\ Rev.\ D }{\bf #1},}
\def\PRL#1,{{\jnfont Phys.\ Rev.\ Lett.\ }{\bf #1},}
\def\MPLA#1,{{\jnfont Mod.\ Phys.\ Lett.\ A }{\bf #1},}
\def\JPG#1,{{\jnfont J.\ Phys.\ G }{\bf #1},}
\def\CTP#1,{{\jnfont Commun.\ Theor.\ Phys.\ }{\bf #1},}
\def\JHEP#1,{{\jnfont JHEP \ }{\bf #1},}
\def\NPPS#1,{{\jnfont Nucl.\ Phys.\ Proc.\ Suppl.\ }{\bf #1},}
\def\CPC#1,{{\jnfont Computl.\ Phys.\ Commun.\ }{\bf #1},}

\def\o_slash{\not{\hbox{\kern-2.1pt $p_1$}}}
\def\p_slash{\not{\hbox{\kern-4.0pt $p_2$}}}
\def\q_slash{\not{\hbox{\kern-2.1pt $q_2$}}}
\def\e_slash{\not{\hbox{\kern-1.5pt $\epsilon_1^*$}}}
\def\f_slash{\not{\hbox{\kern-1.5pt $\epsilon_2^*$}}}

\newcommand{\be}{\begin{equation}}
\newcommand{\ee}{\end{equation}}
\newcommand{\bea}{\begin{eqnarray}}
\newcommand{\eea}{\end{eqnarray}}
\begin{document}

\preprint{\parbox{1.2in}{\noindent ~}}

\title{\ \\[10mm] $R_b$ Constraints on Littlest Higgs Model
                   with T-parity }

\author{Xiao-Fang Han}

\affiliation{Key Laboratory of Frontiers in Theoretical
             Physics,Institute of Theoretical Physics, Academia Sinica,
             Beijing 100190, China \vspace*{1.5cm}}

\begin{abstract}
In the framework of the littlest Higgs model with T-parity (LHT), we
study the contributions of the T-even and T-odd particles to the
branching ratio $R_b$. We find that the precision data
of $R_b$ can give strong constraints on the masses of T-odd fermions.
\vspace*{1cm}
\end{abstract}


\maketitle

\section{Introduction}

The little Higgs theory was proposed \cite{ref1} as a possible
solution to the hierarchy problem and so far remains a popular
candidate for new physics beyond the SM. The littlest Higgs model
\cite{ref2} is a cute economical implementation of little Higgs, but
is found to be subject to strong constraints from electroweak
precision tests \cite{ref3}, which would require raising the mass
scale of the new particles to far above TeV scale and thus
reintroduce the fine-tuning in the Higgs potential \cite{ref4}. To
tackle this problem, a discrete symmetry called T-parity is proposed
\cite{ref5}, which forbids the tree-level contributions from the
heavy gauge bosons to the observables involving only SM particles as
external states. With the running of the LHC, these little Higgs
models will soon be put to the test. Since these little Higgs models
mainly alter the properties of the Higgs boson and the top quark,
hints of these models may be unraveled from various Higgs boson and
top quark processes \cite{higgs-top-lht}.

The branching ratio $R_b$ is defined as
\be
 R_b\equiv\frac{\Gamma(Z\rightarrow b\bar{b})}{\Gamma(Z\rightarrow hadrons)},
\ee
which can provide a precision test of the SM and a sensitive probe
of new physics \cite{np}. In the SM most of
the electroweak oblique and QCD corrections cancel between numerator
and denominator, and the non-decoupling top quark loop effects in the
$Zb\bar{b}$ vertex offer a possibility of bounding the top quark mass.
In the LHT there are new heavy mirror quarks interacting with gauge bosons,
which can contribute to the $R_b$. Therefore, it is possible to give
some constraints on the mirror quark masses via their
radiative corrections to $R_b$.

The contributions of the LHT to $R_b$ was firstly
discussed in \cite{ewlht}, which, however, only considered the
contributions from the diagrams involving the exchange of the SM
Goldstone boson $\pi^\pm$ and neglected the mirror quark contributions
under the assumption of flavor-diagonal and flavor-independent mirror
quark Yukawa couplings. In this paper, we consider the general situation
and examine the contributions of both T-even and T-odd particles to the
$R_b$.

The work is organized as follows. In Sec. II we recapitulate the LHT
model and discuss the new flavor interactions which will contribute
to the decay $Z \to b\bar{b}$. In Sec. III we calculate the one-loop
contributions of the LHT to the branching ratio $R_b$ and present
constraint of $R_b$ on the mirror quark masses. Finally, we give our
conclusions in Sec. IV.

\section{The littlest Higgs model with T-parity}

The LHT model \cite{ref5} is based on a non-linear sigma model
describing the spontaneous breaking of a global $SU(5)$ down to a
global $SO(5)$ by a 5$\times$5 symmetric tensor at the scale
$f\sim\ord({\rm TeV})$. From the $SU(5)/SO(5)$ breaking, there arise
14 Goldstone bosons which are described by the "pion" matrix $\Pi$,
given explicitly by \small \be \label{Pi}
 \addtolength{\arraycolsep}{3pt}\renewcommand{\arraystretch}{1.3}
 \Pi=\left(\begin{array}{ccccc}
-\frac{\omega^0}{2}-\frac{\eta}{\sqrt{20}} &
-\frac{\omega^+}{\sqrt{2}} &
  -i\frac{\pi^+}{\sqrt{2}} & -i\phi^{++} & -i\frac{\phi^+}{\sqrt{2}}\\
-\frac{\omega^-}{\sqrt{2}} &
\frac{\omega^0}{2}-\frac{\eta}{\sqrt{20}} &
\frac{v+h+i\pi^0}{2} & -i\frac{\phi^+}{\sqrt{2}} & \frac{-i\phi^0+\phi^P}{\sqrt{2}}\\
i\frac{\pi^-}{\sqrt{2}} & \frac{v+h-i\pi^0}{2} &\sqrt{4/5}\eta &
-i\frac{\pi^+}{\sqrt{2}} & \frac{v+h+i\pi^0}{2}\\
i\phi^{--} & i\frac{\phi^-}{\sqrt{2}} & i\frac{\pi^-}{\sqrt{2}} &
-\frac{\omega^0}{2}-\frac{\eta}{\sqrt{20}} & -\frac{\omega^-}{\sqrt{2}}\\
i\frac{\phi^-}{\sqrt{2}} &  \frac{i\phi^0+\phi^P}{\sqrt{2}} &
\frac{v+h-i\pi^0}{2} & -\frac{\omega^+}{\sqrt{2}} &
\frac{\omega^0}{2}-\frac{\eta}{\sqrt{20}}
\end{array}\right).
\ee
\normalsize
Under T-parity the SM Higgs doublet $ H= \left(-i
\pi^+/ \sqrt{2}, (v+h+i\pi^0)/2 \right)^T$ is T-even while
other fields are T-odd. A subgroup $[SU(2)\times
U(1)]_{1}\times[SU(2)\times U(1)]_{2}$ of the $SU(5)$ is gauged and
at the scale $f$ it is broken into the SM electroweak symmetry
$SU(2)_L\times U(1)_Y$. The Goldstone bosons $\omega^{0}$,
$\omega^{\pm}$ and $\eta$ are respectively eaten by the new T-odd
gauge bosons $Z_{H}$, $W_{H}$ and $A_{H}$, which obtain masses at
$\ord(v^2/f^2)$ \be M_{W_H}=
M_{Z_H}=fg\left(1-\frac{v^2}{8f^2}\right), ~~ M_{A_H}=\frac{f
g'}{\sqrt{5}}\left(1-\frac{5v^2}{8f^2}\right), \ee with $g$ and
$g^\prime$ being the SM $SU(2)$ and $U(1)$ gauge couplings,
respectively.

The  Goldstone bosons $\pi^{0}$ and $\pi^{\pm}$ are eaten by the
T-even $Z$ and $W$ bosons of the SM, which obtain masses at $\ord(v^2/f^2)$
\be
M_{W_L}=\frac{gv}{2}\left(1-\frac{v^2}{12f^2}\right),\quad
M_{Z_L}=\frac{gv}{2\cos\theta_W}\left(1-\frac{v^2}{12f^2}\right).
\ee
The photon $A_L$ is also T-even and remains massless.

For each SM quark, a copy of mirror quark with T-odd quantum number
is added in order to preserve the T-parity. We denote them by
$u_H^i$ and $d_H^i$, where $i=1, 2, 3$ are the generation index.
In $\ord(v^2/f^2)$ their masses are given by
\be
m_{d_H^i}=\sqrt{2}\kappa_{q^i}f,\qquad
m_{u_H^i}=m_{d_H^i}(1-\frac{v^2}{8f^2}), \label{eq4}
\ee
where $\kappa_{q^i}$ are the diagonalized Yukawa couplings of the mirror
quarks.

Note that new flavor interactions arise between the mirror fermions
and the SM fermions, mediated by the T-odd gauge bosons or T-odd
Goldstone bosons. In general, besides the charged-current
flavor-changing interactions, the FCNC interactions between the
mirror fermions and the SM fermions can also arise from the mismatch
of rotation matrices. For example, there exist FCNC interactions
between the mirror up-type (down-type) quarks and the SM up-type
(down-type) quarks, where the mismatched mixing matrix is denoted by
$V_{H_{u}}$ ($V_{H_{d}}$) with $V^{\dag}_{H_{u}}V_{H_{d}}=V_{CKM}$.
We follow \cite{new-add} to parameterize $V_{H_{d}}$ with three
angles $\theta_{12}^d,\theta_{23}^d,\theta_{13}^d$ and three phases
$\delta_{12}^d,\delta_{23}^d,\delta_{13}^d$ \small \be
\left(\begin{array}{ccc} c_{12}^d c_{13}^d & s_{12}^d c_{13}^d
e^{-i\delta^d_{12}}& s_{13}^d e^{-i\delta^d_{13}}\\
-s_{12}^d c_{23}^d e^{i\delta^d_{12}}-c_{12}^d s_{23}^ds_{13}^d
e^{i(\delta^d_{13}-\delta^d_{23})} & c_{12}^d c_{23}^d-s_{12}^d
s_{23}^ds_{13}^d e^{i(\delta^d_{13}-\delta^d_{12}-\delta^d_{23})} &
s_{23}^dc_{13}^d e^{-i\delta^d_{23}}\\
s_{12}^d s_{23}^d e^{i(\delta^d_{12}+\delta^d_{23})}-c_{12}^d
c_{23}^ds_{13}^d e^{i\delta^d_{13}} & -c_{12}^d s_{23}^d
e^{i\delta^d_{23}}-s_{12}^d c_{23}^d s_{13}^d
e^{i(\delta^d_{13}-\delta^d_{12})} & c_{23}^d c_{13}^d
\end{array}\right).
\ee
\normalsize

\section{$R_b$ in the LHT model}
\begin{figure}[htb]
 \epsfig{file=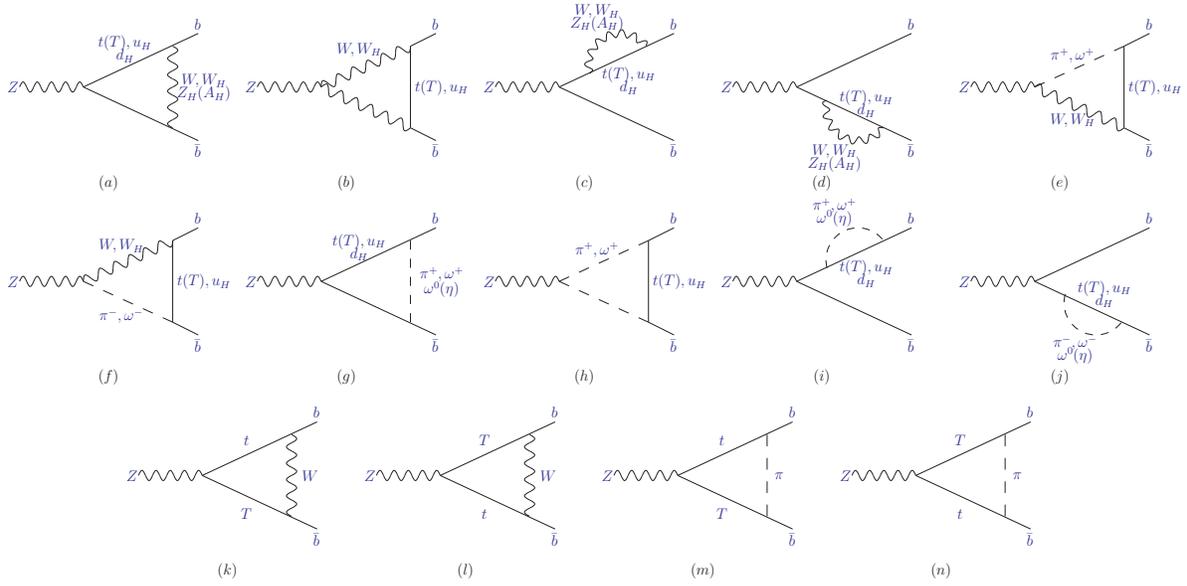,width=16cm}
\vspace*{-0.5cm}
 \caption{Feynman diagrams of $Z \to b\bar{b}$ at one-loop level
          in the LHT model.}
\label{fig1}
 \end{figure}

Fig. \ref{fig1} shows the Feynman diagrams via which LHT gives the
corrections to $\Gamma(Z\to b\bar{b})$. The corrections are from
both T-even and T-odd particles. The contributions of T-even
particles are from the modified coupling $Zt\bar{t}$, $Wt\bar{b}$
and $\pi^{+}t\bar{b}$, and loops involving the top quark T-even
partner (T-quark). The diagrams of T-odd particles are induced by
the interactions between the SM quarks and the mirror quarks
mediated by the heavy T-odd gauge bosons or Goldstone bosons. The
corrections of LHT to the $\Gamma(Z\to d\bar{d})$ and $\Gamma(Z\to
s\bar{s})$ are similar to $\Gamma(Z\to b\bar{b})$ . For the
$\Gamma(Z\to u\bar{u})$ and $\Gamma(Z\to c\bar{c})$, the corrections
are only from the T-odd particles, and corrections from the T-even
particle can be neglected safely due to the small coupling of
$ZT\bar{u}$ and $ZT\bar{c}$. In this work, our purpose is to examine
the $R_b$ dependence on the mirror quarks mass, and adopt the method
of Bernabeu, Pich, and Santamaria (BPS) to calculate various
hadronic decay widths of $Z$ boson \cite{bps1,bps2,rbnp}. In
Appendix, we present the calculation in detail.

In LHT, the branching ratio of $Z\to b\bar{b}$ can be expressed as
 \be R_b \simeq R_b^{SM}(1+\frac{\delta\Gamma_b}
 {\Gamma_b^{SM}}-R_b^{SM}
 \frac{\delta\Gamma_{had}}{\Gamma_b^{SM}}),
 \label{eq5} \ee
where $R_b^{SM}$ and $\Gamma_b^{SM}$ are the SM predictions for the
branching ratio of $Z\to b\bar{b}$ and the width $\Gamma(Z\to
b\bar{b})$, $\delta\Gamma_b$ and $\delta\Gamma_{had}$ are the
correction of LHT to the $\Gamma_b^{SM}$ and $\Gamma^{SM}(Z\to
hadrons)$, respectively.

In the numerical calculations we take the Fermi constant $G_F$ , the
fine-structure constant $\alpha_{M_{Z_L}}$, $Z$-boson mass
$M_{Z_L}$, fermion masses $m_f$, and the electroweak mixing angle
$s_W = \sin\theta_W$ as input parameters \cite{pdg}. The LHT
parameters relevant to our calculation are the scale $f$, the ratio
between top quark Yukawa couplings $r=\frac{\lambda_1}{\lambda_2}$,
the mirror quark masses and parameters in the matrices $V_{H_u}$ and
$V_{H_d}$. $f$ may be as low as 500 GeV \cite{ewlht}, and $r$ is
taken typical value as 1. For the mirror quark masses, from
Eq.(\ref{eq4}) we get $m_{u^i_{H}}=m_{d^i_{H}}$ at $\ord(v/f)$ and
further we assume \be
m_{u^1_{H}}=m_{u^2_{H}}=m_{d^1_{H}}=m_{d^2_{H}}\equiv M_{12},\ \
m_{u^3_H}=m_{d^3_H}\equiv M_3. \ee For  the matrices $V_{H_u}$ and
$V_{H_d}$, considering the constraints in \cite{flalht}, we follow
them to consider the following four scenarios:
\begin{itemize}
\item[(I)]  $V_{H_u}=1$, $V_{H_d}=V_{CKM}$.
\item[(II)]  $V_{H_d}=1$, $V_{H_u}=V_{CKM}^{\dagger}$.
\item[(III)]  $s^d_{13}=0.5$, $\delta^d_{12}=\delta^d_{23}=0$,
              $\delta^d_{13}=\delta^{SM}_{13}$,
              $s^d_{ij}=s^{SM}_{ij}$ otherwise.
\item[(IV)]  $s^d_{13}=0.5$, $s^d_{12}=0.7$, $s^d_{23}=0.4$,
             $\delta^d_{12}=\delta^d_{23}=0$,
             $\delta^d_{13}=\delta^{SM}_{13}$.
 \end{itemize}
\begin{figure}[htb]
 \epsfig{file=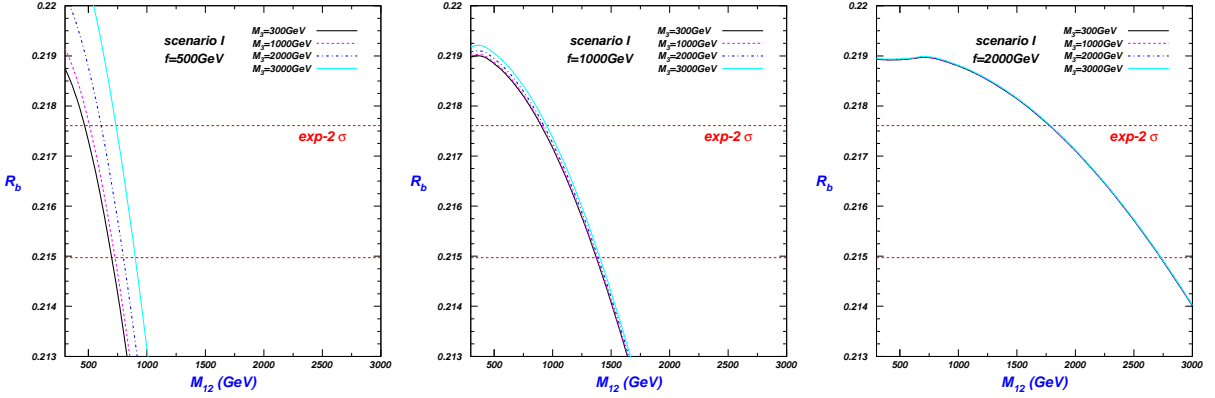,width=16cm}
\vspace*{-0.5cm}
 \caption{The branching ratio $R_b$ versus the mass of first two family mirror
 quarks in scenario-I with $f=500$ GeV, $1000$ GeV and $2000$ GeV, respectively.}
 \label{fig2}
 \end{figure}
\begin{figure}[htb]
 \epsfig{file=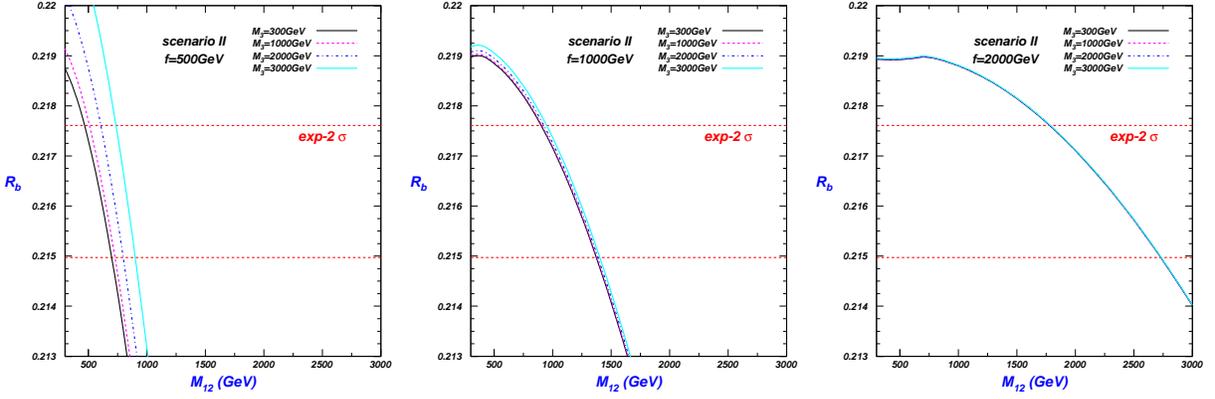,width=16cm}
\vspace*{-0.5cm}
 \caption{Same as Fig. 2, but for scenario-II.}
\label{fig3}
 \end{figure}

In Figs. 2-5, we plot the branching ratio $R_b$ versus the first two
mirror quark mass $M_{12}$ for the scenario I, II, III and IV,
respectively. The Figs. 2-5 show  $R_b$ can give strict lower bound
and upper bound of the first two mirror quark mass for the $f$ and
$M_3$ taken. The constraints are sensitive to the scale $f$, and the
allowed regions of $M_{12}$ become larger with the increasing $f$.
Further, the $R_b$ favors a large value of $M_{12}$ for a large
value of $f$.

In scenario I and scenario II, the up-type Yukawa interactions and
the down-type quark Yukawa interactions are diagonal, respectively.
However, scenario III and scenario IV are two large mixing
scenarios, and the angle $s^d_{13}$ is set large so that the third
generation mass dependence can be more sensitive. For example, when
$f=1$ TeV ($2$ TeV), the four lines in Fig. 2 and Fig. 3 are almost
overlapped for scenario I and scenario II, and this situation can be
relaxed for scenario III and scenario IV. Besides, for $f=500$ GeV,
$M_3=3000$ GeV can be allowed in scenario I and scenario II, but be
ruled out in scenario III and scenario IV by the 2$\sigma$ $R_b$
constraints.

\begin{figure}[htb]
 \epsfig{file=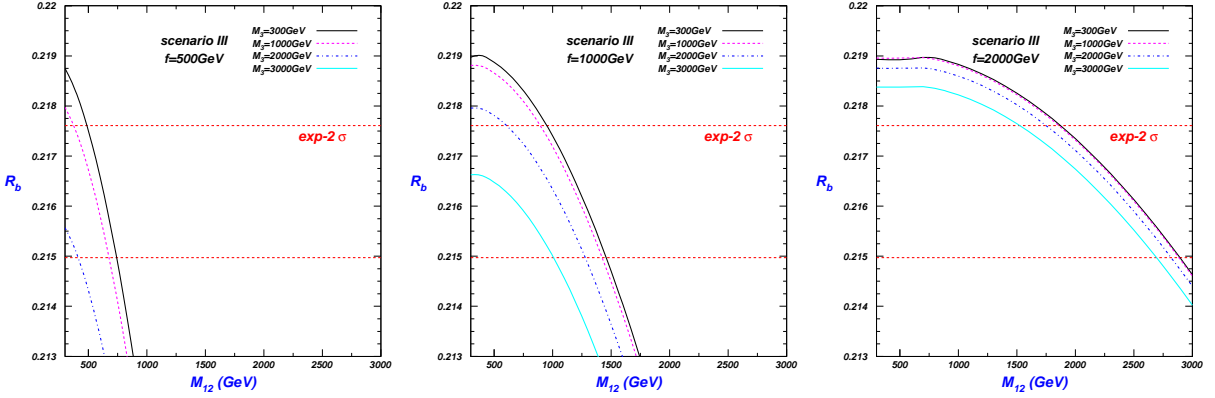,width=16cm}
\vspace*{-0.5cm}
 \caption{Same as Fig. 2, but for scenario-III.}
\label{fig4}
 \end{figure}
\begin{figure}[htb]
 \epsfig{file=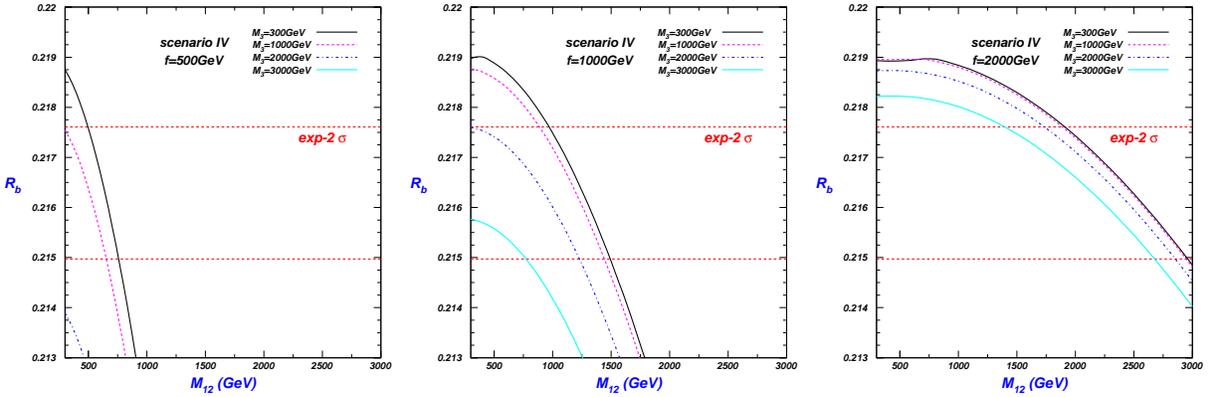,width=16cm}
\vspace*{-0.5cm}
 \caption{Same as Fig. 2, but for scenario-IV.}
\label{fig5}
 \end{figure}

\section{Conclusion}
In the framework of littlest Higgs model with T-parity, we studied the
loop contributions of the T-even and T-odd particles to the branching
ratio $R_b$ for four different scenarios. We found that the
precision measurement data of $R_b$ can give strong constraints on
the mirror quark masses. For the values of  $f$ and $M_3$ in various
scenarios of $V_{H_d}$, $R_b$ can give strict lower bound and upper
bound for the mass $M_{12}$ of the first two generations of mirror quarks,
and the allowed regions of $M_{12}$ become larger as $f$ gets large. Further,
the $R_b$ data favors a large value of $M_{12}$ in case of a large $f$.
Besides, the $R_b$ constraints on the masses of three
generation mirror quarks depend on the texture of $V_{H_d}$, and are
more sensitive to the mass $M_3$ of the third generation of mirror quarks
in scenarios III and IV than in scenarios I and II.
For example, when $f=500$ GeV, $M_3=3000$ GeV is allowed in scenarios I and
II, but ruled out in scenarios III and IV by the 2$\sigma$ $R_b$ constraints.

\section*{Acknowledgement}
We thank L. Wang, J. M. Yang and C. P. Yuan for discussions.
This work was supported  by the National Natural
Science Foundation of China (NNSFC) under Nos. 10821504,
10725526 and 10635030.

\appendix
\section{The heavy quark loop contributions to $Z \to b\bar{b}$}
According to the BPS method \cite{bps1,bps2,rbnp}, we give the
expressions of hadronic decay widths of $Z$-boson:
 \be
 \Gamma_q=\frac{3m_Z}{12\pi}(v_q^2+a_q^2)[1+\frac{\alpha}{\pi }(\frac{Z^{q}_{L}}
{(Z^{q}_{L})^2+(Z^{q}_{R})^2}) F_q] ~~(q=b,~d,~s,~u,~c),
\label{app4}
\ee
where $v_q=\frac{Z_L^q+Z_R^q}{2}$ and $a_q=\frac{Z_L^q-Z_R^q}{2}$
with $Z_L^q$ and $Z_R^q$ being the
left- and right-handed couplings of $Zq\bar{q}$, respectively.
\bea
F_{b,d,s}=&&V_{cha}(t,W,\pi)+V_{cha}(T,W,\pi)+V_{cha}(u_H^i,W_H,\omega)+
V_{neu}(d_H^i,Z_H,\omega^0)\nonumber \\
&&+V_{neu}(d_H^i,A_H,\eta)+V_{mix}(t,T,W,\pi),\nonumber\\
F_{u,c}=&&V_{cha}(d_H^i,W_H,\omega)+ V_{neu}(u_H^i,Z_H,\omega^0)+V_{neu}(u_H^i,A_H,\eta),
\eea
where
\begin{eqnarray}
&&V_{cha}(f,V,S)=F^{(a)}+F^{(b)}+F^{(c)+(d)}+F^{(e)+(f)}+F^{(g)}+F^{(h)}+F^{(i)+(j)}, \nonumber\\
&&V_{neu}(f,V,S)=F^{(a)}+F^{(c)+(d)}+F^{(g)}+F^{(i)+(j)}, \nonumber\\
&&V_{mix}(f,f',V,S)=F^{(k)+(l)}+F^{(m)+(n)}.
\end{eqnarray}
$F^{(a)-(n)}$ in the above equations are the corresponding explicit
expressions of the Feynman diagrams in Fig. \ref{fig1}, which are
given by
\begin{eqnarray}
&&F^{(a)}=-\frac{1}{g^2 s_w^2}|c_3^f|^2
\big\{\frac{Z_L^f}{2}\big[\frac{r(r-2)}{(r-1)^2}\ln
r+\frac{r}{r-1}\big]+Z_R^f\big[\frac{r}{(r-1)^2}\ln
r-\frac{r}{r-1}\big]\big\}, \nonumber\\
&&F^{(b)}=-\frac{3}{2g^2s_w^2}|c_3^f|^2
g_{ZVV}\big[\frac{r^2}{(r-1)^2}\ln
r-\frac{r}{r-1}\big], \nonumber\\
&&F^{(c)+(d)}=\frac{1}{2g^2s_w^2}|c_3^f|^2Z_L^q
\big[\frac{r^2}{(r-1)^2}\ln
r-\frac{r}{r-1}\big], \nonumber\\
&&F^{(e)+(f)}=\frac{1}{g^2s_w^2}c_3^f a_3^{f*}
g_{ZVS}\big[\frac{r}{(r-1)^2}\ln
r-\frac{1}{r-1}\big]\frac{1}{m_V}, \nonumber\\
&&F^{(g)}=-\frac{1}{2g^2s_w^2r}|a_3^f|^2
\big\{\frac{Z_R^f}{2}\big[\Delta+\frac{r(r-2)}{(r-1)^2}\ln
r+\frac{2r-1}{r-1}\big]+Z_L^f\big[\frac{r}{(r-1)^2}\ln
r-\frac{r}{r-1}\big]\big\}, \nonumber\\
&&F^{(h)}=\frac{1}{4g^2s_w^2r}|a_3^f|^2g_{ZSS}\big[\Delta+\frac{r^2}{(r-1)^2}\ln
r-\frac{r}{r-1}\big], \nonumber\\
&&F^{(i)+(j)}=\frac{1}{4g^2s_w^2r}|a_3^f|^2Z_L^q\big[\Delta+\frac{r^2}{(r-1)^2}\ln
r-\frac{r}{r-1}\big], \nonumber\\
&&F^{(k)+(l)}=-\frac{2}{g^2s_w^2}c_3^t c_3^{T*}
\big\{\frac{Z_L^{tT}}{2}\frac{1}{r'-r}\big[\frac{r'^2}{r'-1}\ln
r'-\frac{r^2}{r-1}\ln r\big]
\nonumber\\
&&~~~~~~~~~~~~~-Z_R^{tT}\sqrt{rr'}\frac{1}{r'-r}\big[\frac{r'}{r'-1}\ln
r'-\frac{r}{r-1}\ln r\big]\big\}, \nonumber\\
&&F^{(m)+(n)}=\frac{1}{2g^2s_w^2}a_3^t
a_3^{T*}\big\{\frac{2Z_L^{tT}}{r'-r}\big[\frac{r'}{r'-1}\ln
r'-\frac{r}{r-1}\ln r
\big]\nonumber\\
&&~~~~~~~~~~~~~~-\frac{Z_R^{tT}}{\sqrt{rr'}}(\Delta+1+\frac{1}{r'-r}\big[\frac{r'^2}{r'-1}\ln
r'-\frac{r^2}{r-1}\ln r \big] ) \big\},
\end{eqnarray}
with
\be
\Delta\equiv\frac{2}{n-4}+\gamma+\ln (m_V^2/4\pi\mu^2)-\frac{3}{2},
\ee
$r=m_f^2/m_V^2$, $r'=m_{f'}^{2}/m_V^2$. The coupling constant appearing above are
from
\begin{eqnarray*}
V\bar{f}q&:&i\gamma^{\mu}(c_3^f P_L+d_3^f P_R),\hspace{2.5cm}
S\bar{f}q:a_3^f P_L+b_3^f P_R,\nonumber\\
Z\bar{f}f&:&i\gamma^{\mu}(Z_L^f P_L+Z_R^f P_R),\hspace{2.15cm}
Z\bar{t}T:i\gamma^{\mu}(Z_L^{tT} P_L+Z_R^{tT} P_R),\nonumber\\
ZS^+S^-&:&ig_{VSS} (p_{S^+}^{\mu}-p_{S^-}^{\mu}),\hspace{1.8cm}
ZV^+ S^-:g_{ZVS}g^{\mu\nu},\nonumber\\
Z^\rho V^{+\mu} V^{-\nu}&:&ig_{ZVV}[g^{\mu\nu}(p_{+}-p_{-})^{\rho}
+g^{\nu\rho}(p_{-}-p_{Z})^{\mu}+g^{\rho\mu}(p_{Z}-p_{+})^{\nu}],
\end{eqnarray*}
where $f$, $V$ and $S$ represent fermion, gauge bosons and scalar
particles involved in the loops, respectively. The explicit
expressions of these parameters are complicated at $\ord(v^2/f^2)$
and can be found in \cite{blht,zsblht}.

\end{document}